\documentclass[preprint]{aastex61}
\usepackage{amssymb,amsmath}
\usepackage{graphicx}
\usepackage{natbib}
\usepackage[OT1,T1]{fontenc}

\begin{document}

\title{Non-LTE Calculations of the \ion{Fe}{1} 6173 \AA\ Line in a Flaring Atmosphere}

\author{Jie Hong}
\affiliation{School of Astronomy and Space Science, Nanjing University, Nanjing 210023, China}
\affiliation{Key Laboratory for Modern Astronomy and Astrophysics (Nanjing University), Ministry of Education, Nanjing 210023, China}

\author{M.~D. Ding}
\affiliation{School of Astronomy and Space Science, Nanjing University, Nanjing 210023, China}
\affiliation{Key Laboratory for Modern Astronomy and Astrophysics (Nanjing University), Ministry of Education, Nanjing 210023, China}

\author{Ying Li}
\affiliation{Key Laboratory of Dark Matter and Space Astronomy, Purple Mountain Observatory, Chinese Academy of Sciences, Nanjing 210034, China}

\author{Mats Carlsson}
\affiliation{Rosseland Centre for Solar Physics, University of Oslo, P.O. Box 1029 Blindern, NO-0315 Oslo, Norway}
\affiliation{Institute of Theoretical Astrophysics, University of Oslo, P.O. Box 1029 Blindern, NO-0315 Oslo, Norway}

\email{dmd@nju.edu.cn}

\begin{abstract}
The \ion{Fe}{1} 6173 \AA\ line is widely used in the measurements of vector magnetic fields by instruments including the Helioseismic and Magnetic Imager (HMI). We perform non-local thermodynamic equilibrium calculations of this line based on radiative hydrodynamic simulations in a flaring atmosphere. We employ both a quiet-Sun atmosphere and a penumbral atmosphere as the initial one in our simulations. We find that, in the quiet-Sun atmosphere, the line center is obviously enhanced during an intermediate flare. The enhanced emission is contributed from both radiative backwarming in the photosphere and particle beam heating in the lower chromosphere. A blue asymmetry of the line profile also appears due to an upward mass motion in the lower chromosphere. If we take a penumbral atmosphere as the initial atmosphere, the line has a more significant response to the flare heating, showing a central emission and an obvious asymmetry. The low spectral resolution of HMI would indicate some loss of information but the enhancement and line asymmetry are still kept. By calculating polarized line profiles, we find that the Stokes \textit{I} and \textit{V} profiles can be altered as a result of flare heating. Thus the distortion of this line has a crucial influence on the magnetic field measured from this line, and one should be cautious in interpreting the magnetic transients observed frequently in solar flares.
\end{abstract}

\keywords{line: profiles --- radiative transfer --- Sun: flares --- Sun: photosphere}

\section{Introduction}
The \ion{Fe}{1} 6173 \AA\ line is thought to be formed in the photosphere, with a formation height of around 200--300 km \citep{2006norton,2009bello}. This line is magnetically sensitive and has a good performance in vector magnetic field measurements \citep{2006norton}. Thus it has been selected in many solar instruments for polarimetric observations including the Helioseismic and Magnetic Imager (HMI; \cite{2012schou}). In quiet regions and even regions with some activity, this line is mostly in absorption and quite stable. However, it is possible that this line can be affected by flares, in which the deeper layers are more or less heated. The change in the shape of the line profile has an influence on the inversions made to determine the magnetic field.

One example is the magnetic transients whose origin is still under  debate. They are observed as sudden changes in the magnetic field and even in the sign in some flares \citep{1981patterson,1984patterson}. \cite{1984patterson} argued that flare emissions can cause magnetic transients. \cite{2003qiu} found that magnetic transients occur at the location of strong hard X-ray emissions. Many other studies also reported such magnetic transients that are not ``real'', but results of the altered line profiles during flares \citep{2009maurya,2012maurya}.

Previous studies on magnetic transients mainly employ data from the Michelson Doppler Imager \citep{1995scherrer}, which uses the \ion{Ni}{1} 6768 \AA\ line. \cite{2002ding} carried out a detailed study of this line in a flaring atmosphere and found that this line can go into emission when a relatively cool atmosphere (like the sunspot penumbra) is heated by an electron beam. \cite{2013harker} first used a semi-empirical flare model to synthesize the \ion{Fe}{1} 6173 \AA\ line profile observed by HMI. \cite{2017sharykin} calculated the line profile from RADYN simulation results, and found that the intensity change is still smaller in magnitude compared with observations. Their calculations are only based on the quiet Sun atmosphere. However, many magnetic transients are observed in umbral and/or penumbral regions \citep{2015burtseva,2001kosovichev}. Therefore, we should also consider a cooler initial atmosphere as in \cite{2002ding}.

There has been a large number of magnetic transient observations with HMI, yet the detailed process still needs more investigation, especially the response to flare heating. In this Letter, we investigate the response of the \ion{Fe}{1} 6173 \AA\ line to flare heating for both a quiet-Sun atmosphere and a penumbral atmosphere using radiative hydrodynamic simulations. The remainder of this Letter is arranged as follows. In Chapter 2 we describe the basic method and model parameters. We present the results for the two different atmospheres in Chapter 3, followed by a summary and discussions in Chapter 4.

\section{Method}
RADYN is a radiative hydrodynamics code \citep{1992carlsson,1995carlsson,1997carlsson,2002carlsson} that can implicitly solve the hydrodynamic and radiative transfer equations together using an adaptive grid \citep{1987dorfi}. It has later been applied to flare simulations \citep{1999abbett,2005allred,2015allred} to calculate the atmospheric response to a beam of non-thermal electrons injected at the coronal loop top. The electron beam heating rate is obtained by solving the Fokker-Planck equation following \cite{1990mctiernan}. Atoms that are important in the chromosphere are treated in non-LTE: a hydrogen atom with six levels including continuum, a \ion{Ca}{2} atom with six levels including continuum, and a helium atom with nine levels including continuum. All radiative transitions between the energy levels mentioned above are treated in complete frequency redistribution. Other atoms are treated in LTE with the background opacity package of \cite{1973gustafsson}.

\begin{figure}
\plotone{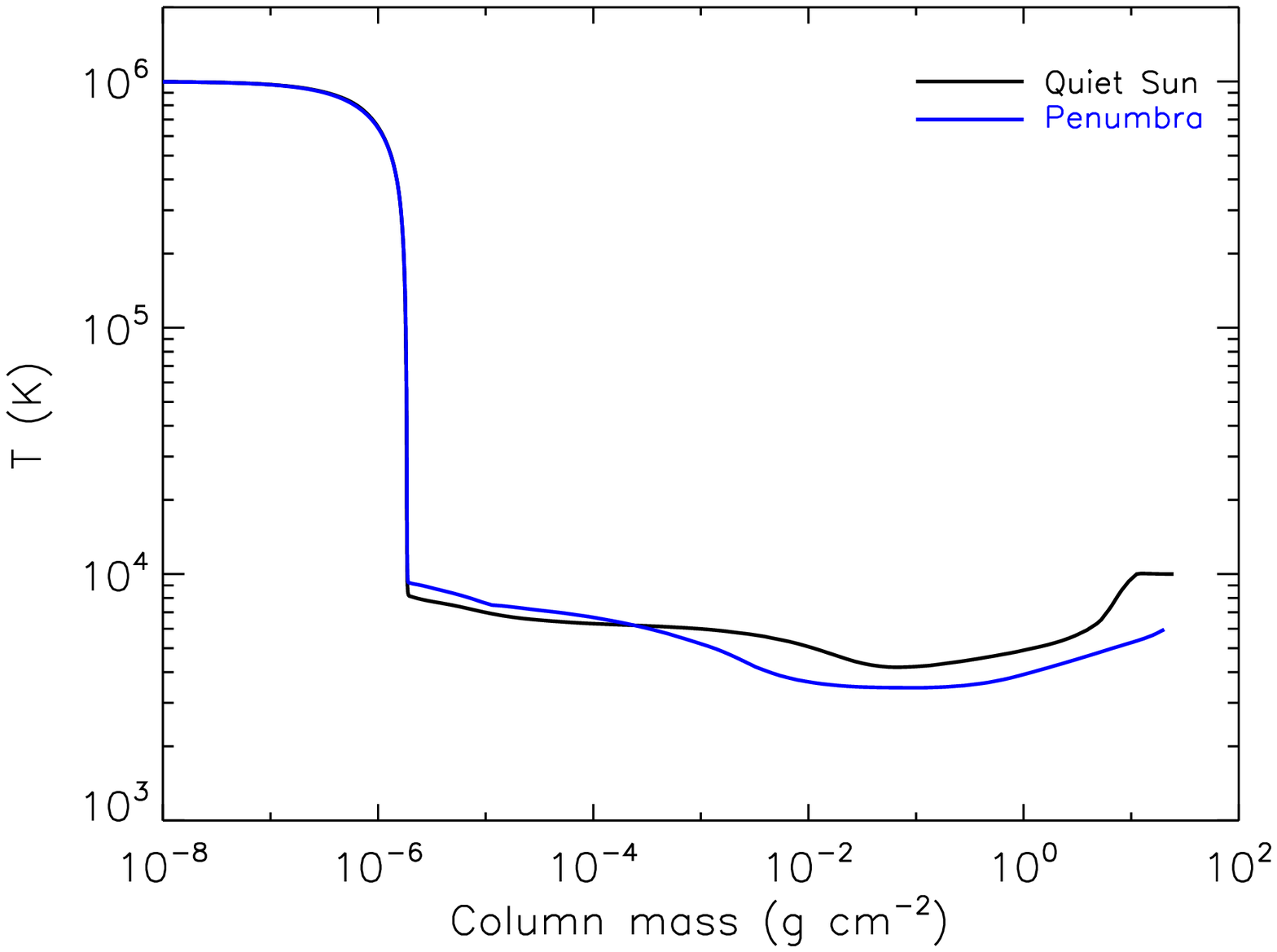}
\caption{Distribution of temperature of the initial atmosphere for RADYN simulations after relaxation. The black curve is for the quiet Sun and the blue curve is for the penumbra, respectively.}
\label{atm}
\end{figure}

We assume an atmosphere in a 10 Mm quarter-circular loop structure. Following \cite{2017hong}, we employ two different atmospheric models to construct the initial atmosphere. One is a quiet-Sun atmosphere, the same as in \cite{2017hong}, which is based on the VAL3C model \citep{1981vernazza}; the other is a penumbral atmosphere that is based on the semi-empirical model of \cite{1989ding}. Fig.~\ref{atm} shows the two initial model atmospheres after relaxation, in which the penumbral model has a lower temperature in the lower atmosphere.

We assume that the initial atmosphere is subject to an electron beam heating during a flare. The electron beam is assumed to have a power-law distribution with a spectral index of 3. The energy flux of the beam follows a triangular function over time, with a total duration of 20 s. We consider two cases here: one with an average energy flux (half of the maximum; $F$) of $5\times 10^{10}$ erg cm$^{-2}$ s$^{-1}$ and an ordinary low-energy cutoff ($E_{c}$) of 25 keV; the other with a lower average energy flux of $5\times 10^{9}$ erg cm$^{-2}$ s$^{-1}$ and a much higher low-energy cutoff of 200 keV. The former represents an intermediate flare. The latter is to mimic an extreme case like the white-light flare reported by \cite{2006xu} in which the electrons can be accelerated to very high energies. We run both cases with the two different initial atmospheres respectively. The model parameters of all four running cases are listed in Table~\ref{tab}.

Each case is run for 20 s and we save the simulation snapshots every 0.1 s. The snapshots are then taken as input to the radiative transfer code RH \citep{2001uitenbroek,2015pereira} to calculate the \ion{Fe}{1} 6173 \AA\ line profiles. For better convergence, when calculating this line, we cut the atmosphere above 1.2 Mm where the temperature is relatively high and the contribution to this line is negligible.

\begin{table}[ht]
\centering
\caption{List of parameters of flare simulation}
\begin{tabular}{cccccc}
\hline
Label & $F$ (erg cm$^{-2}$ s$^{-1}$) & Total duration (s) & Spectral index & $E_{c}$ (keV) & Initial Atmosphere\\
\hline
FQa & $5\times 10^{10}$ & 20  & 3 & 25 & Quiet Sun\\
FQb & $5\times 10^{9}$ & 20  & 3 & 200 & Quiet Sun\\
FPa & $5\times 10^{10}$ & 20  & 3 & 25 & Penumbra\\
FPb & $5\times 10^{9}$ & 20  & 3 & 200 & Penumbra\\
\hline
\end{tabular}
\label{tab}
\end{table}

\section{Results}
\subsection{Quiet-Sun atmosphere}
We show the evolution of the \ion{Fe}{1} 6173 \AA\ line profile for all the four cases in Fig.~\ref{line}. For cases FQa and FQb, we see a dimming in both the line and the nearby continuum during the first 2 s. The dimming effect caused by non-thermal electrons seems a common phenomenon in flares and Ellerman bombs, which has already been reported in some previous simulations \citep{1999abbett,2006allred,2017hong}. Dimming in this line is also seen in the calculation results of \cite{2017sharykin}. At the beginning, the line source function decouples from the local Planck function above the temperature minimum region (TMR), and the formation height of the line center (optical depth unity) is around 220 km (Fig.~\ref{flarea}). When flare heating sets in, the line formation height increases first, resulting in a dimming effect, and then decreases gradually when the upper atmosphere is heated. The decoupling between the line source function and the Planck function is also reduced since the non-thermal heating enhances the line source function. As a result, the contribution function to the line extends to a higher level in the atmosphere. At 10 s and 12 s, the contribution function shows a small hump at around 550 km in addition to the main hump at the lower layers. An analysis of the energy budget in Case FQa shows that the electron beam heating is the dominant gain in the chromosphere (above 500 km), while radiative heating (backwarming) is important in the photosphere where the line center is formed. These two energy terms contribute to a significant intensity increase at the line center, although the line profile remains to be in absorption. The line profile shows a very weak blue asymmetry due to an upward velocity ($\sim$ 1 km s$^{-1}$) in the lower chromosphere.

Case FQb adopts a very high cutoff energy of the electron beam, which is used to mimic a larger penetration depth of the beam electrons in some exceptionally energetic white-light flares (e.g. \cite{2006xu}). From Fig.~\ref{flarea} we can see that direct heating by non-thermal electrons can indeed go into deeper layers than in FQa. Although the energy flux of the electron beam is one order smaller of magnitude in Case FQb, the increase at the line center has a similar magnitude to that in Case FQa. Naturally, in Case FQb, the contribution of electron beam heating to the line intensity is relatively larger than in Case FQa, as judged from the obvious hump in the contribution function at 10 s and 12 s. There is still a weak blue asymmetry in the line profile in Case FQb.

In order to compare the model results with real HMI observations, we then convolve our calculated line profiles with the six HMI transmission functions to obtain the simulated HMI profiles following \cite{2013harker}. These results are superimposed to Fig.~\ref{line} as diamonds. It is seen that although with a lower spectral resolution, the simulated HMI profiles still show a clear  intensity increase at the line center due to flare heating. However, since the positions of the six wavelength points are not symmetric to the line center, the weak blue asymmetry, found in some original profiles as mentioned above, is not clearly seen in the simulated HMI profiles. Instead, there appears a fake red asymmetry in the simulated HMI profiles for the quiet-Sun cases considered here.

\subsection{Penumbral atmosphere}
Compared to the quiet Sun, the penumbra has a lower temperature in the lower atmosphere, but the temperature enhancement could be larger if subjected to the same electron beam (Fig.~\ref{flareb}). For Cases FPa and FPb, we do not see any dimming in the line, but a dimming in the continuum (Fig.~\ref{line}). The non-thermal electrons penetrate deeper in the atmosphere than in the quiet-Sun case. One can see a complex profile of the contribution function with two components. At 5 s of Case FPa, the enhancement at the line center is still mainly caused by the backwarming effect at the line formation height, although there exists another small contribution from the beam heating. However, at the same time of Case FPb, the beam heating above the line formation height already plays a major role. In the latter case, the line source function in the chromosphere is greatly enhanced by the electron beam heating. As a result, the place where the contribution function reaches its maximum is higher than the height of optical depth unity. The line center is much more enhanced in the penumbral cases than in the quiet-Sun cases. Although the line profiles are generally in absorption, there appear an emission peak at the line center. However, the emission peak is not visible in the HMI profiles owing to the low spectral resolution of HMI, but the intensity enhancement at the line core is still larger compared with the quiet-Sun cases. We also notice a blue asymmetry in the original calculated line profile, shown as an enhanced blue wing and a weak blue shift of the central emission peak. The blue asymmetry comes from the mass motion in the layers of around 500 km where the contribution function is still large, although these layers are well above the height of optical depth unity. In the simulated HMI profiles, a fake red asymmetry appears at the beginning as in the quiet-Sun cases, and it then gradually turns to a blue asymmetry.

\subsection{Polarized line profiles}
We have also modeled the Stokes line profiles using the RH code, assuming a vertical magnetic field with an exponential distribution $B(z)=B_{0}\textrm{e}^{-z/H}$. The quantity $B_{0}$ is the magnetic field strength at the bottom of the photosphere, which is set to 100 G for the quiet Sun (cases FQa and FQb), and 1000 G for the penumbra (cases FPa and FPb). The scale height $H$ is specifically set so that the magnetic field would reduce to half of $B_{0}$ at 200 km. The results are shown in Fig.~\ref{stokes}. 

\begin{figure}
\plotone{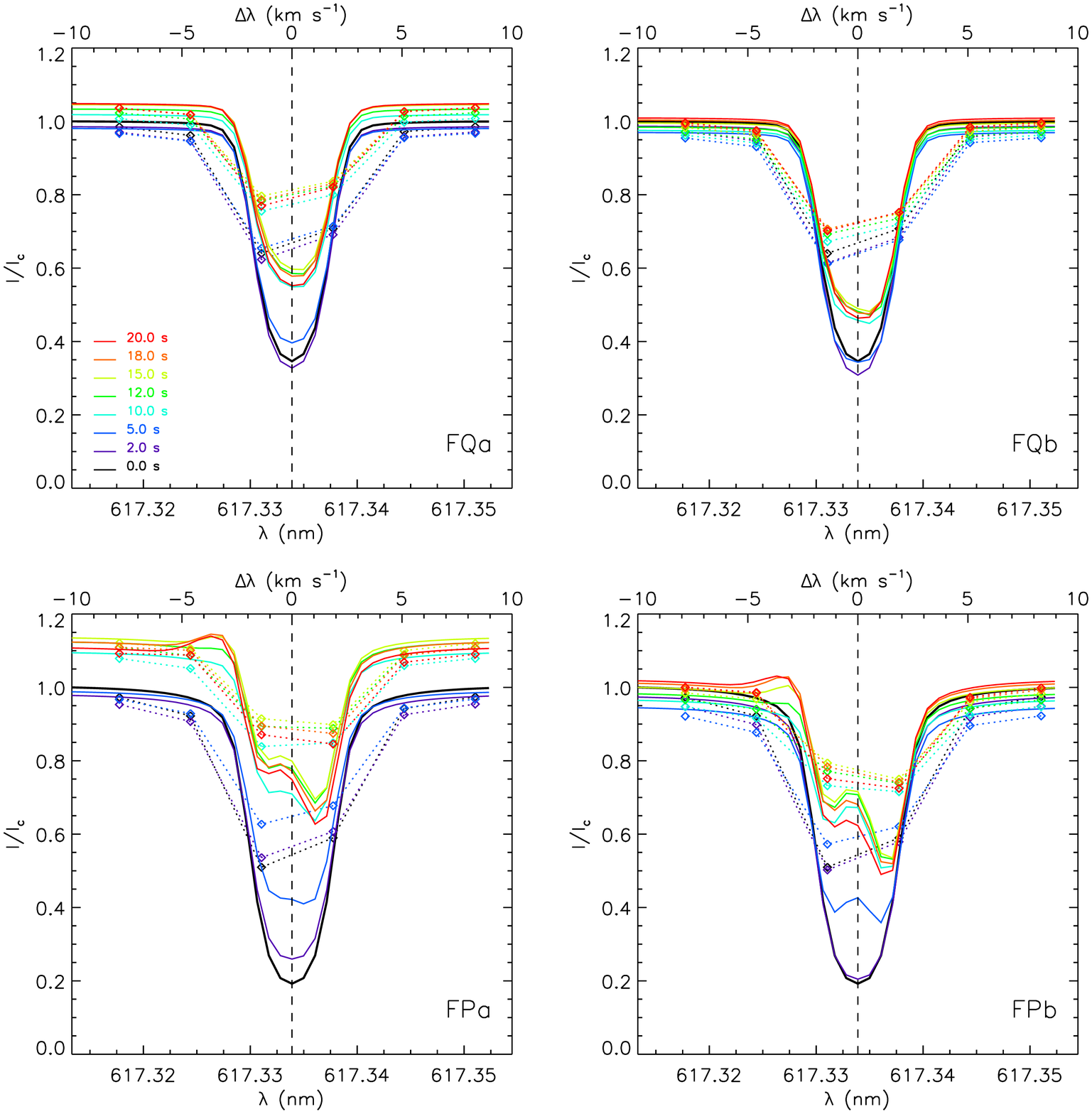}
\caption{Time evolution of the \ion{Fe}{1} 6173 \AA\ line profiles in the four simulation cases. The horizontal axis above marks the wavelength in Doppler velocity units, with negative sign for blueshift and positive sign for redshift. Dashed vertical lines denote the line center wavelength. Synthetic line profiles are shown in solid lines, while the simulated HMI profiles are shown as diamonds representing the observed values with the six tunable filters. }
\label{line}
\end{figure}

\begin{figure}
\plotone{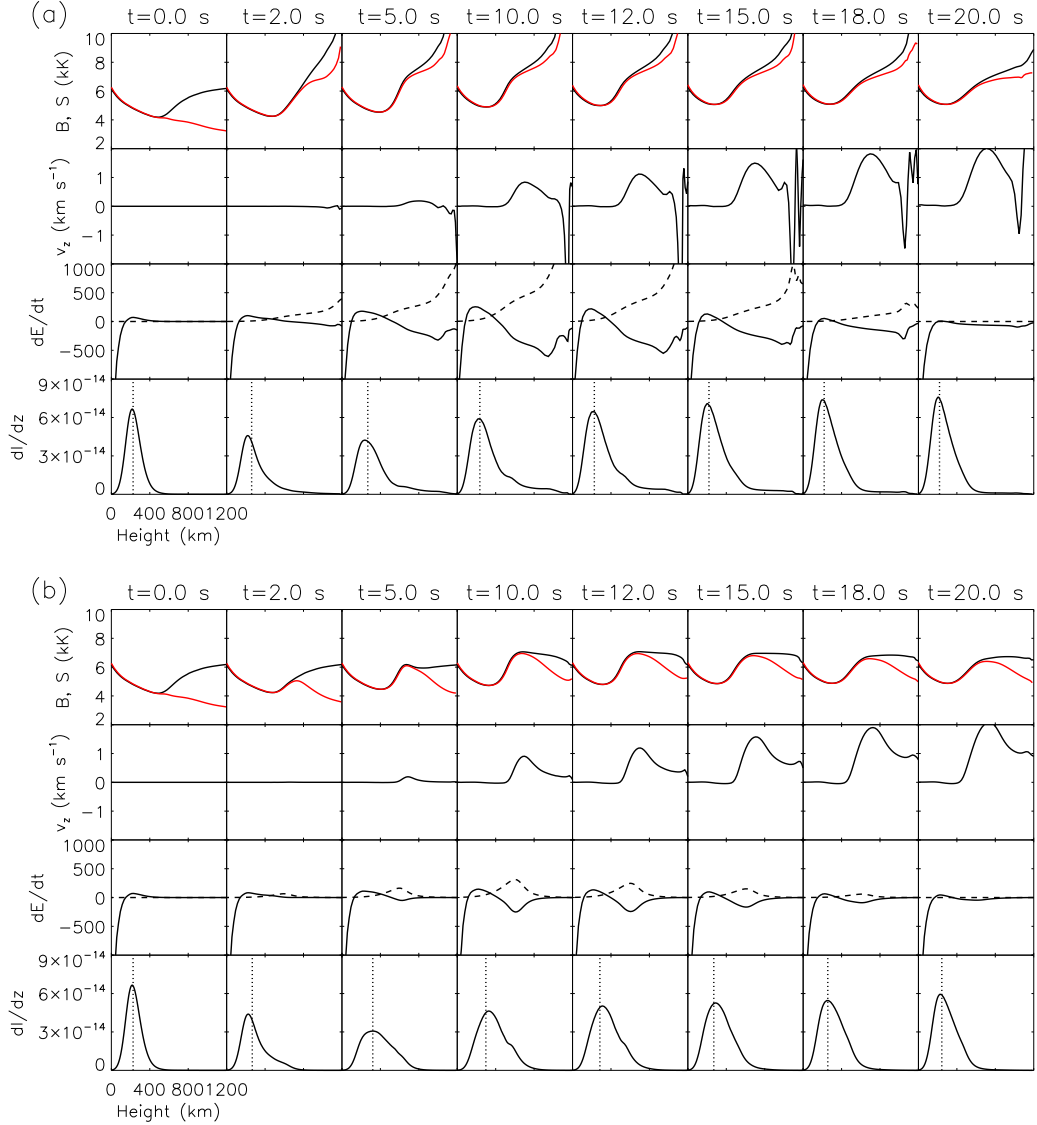}
\caption{Time evolution of the Planck function and the line source function of \ion{Fe}{1} 6173 \AA , the vertical velocity, the heating rates, and the contribution function to the line intensity in Cases FQa (a) and FQb (b). In the top row of each panel, the Planck function (black) and the line source function (red) are plotted as an ``equivalent'' temperature. In the second row of each panel, a positive velocity means an upflow, while a negative one means a downflow, which is opposite to the definition in Fig.~\ref{line}. In the third row of each panel, the solid curve shows the radiative heating rate and the dashed curve shows the electron beam heating rate, with units of erg cm$^{-3}$ s$^{-1}$. In the last row of each panel, the contribution function is shown as the solid curve, while the dotted line denotes the height where the optical depth at the line center reaches unity.}
\label{flarea}
\end{figure}

\begin{figure}
\plotone{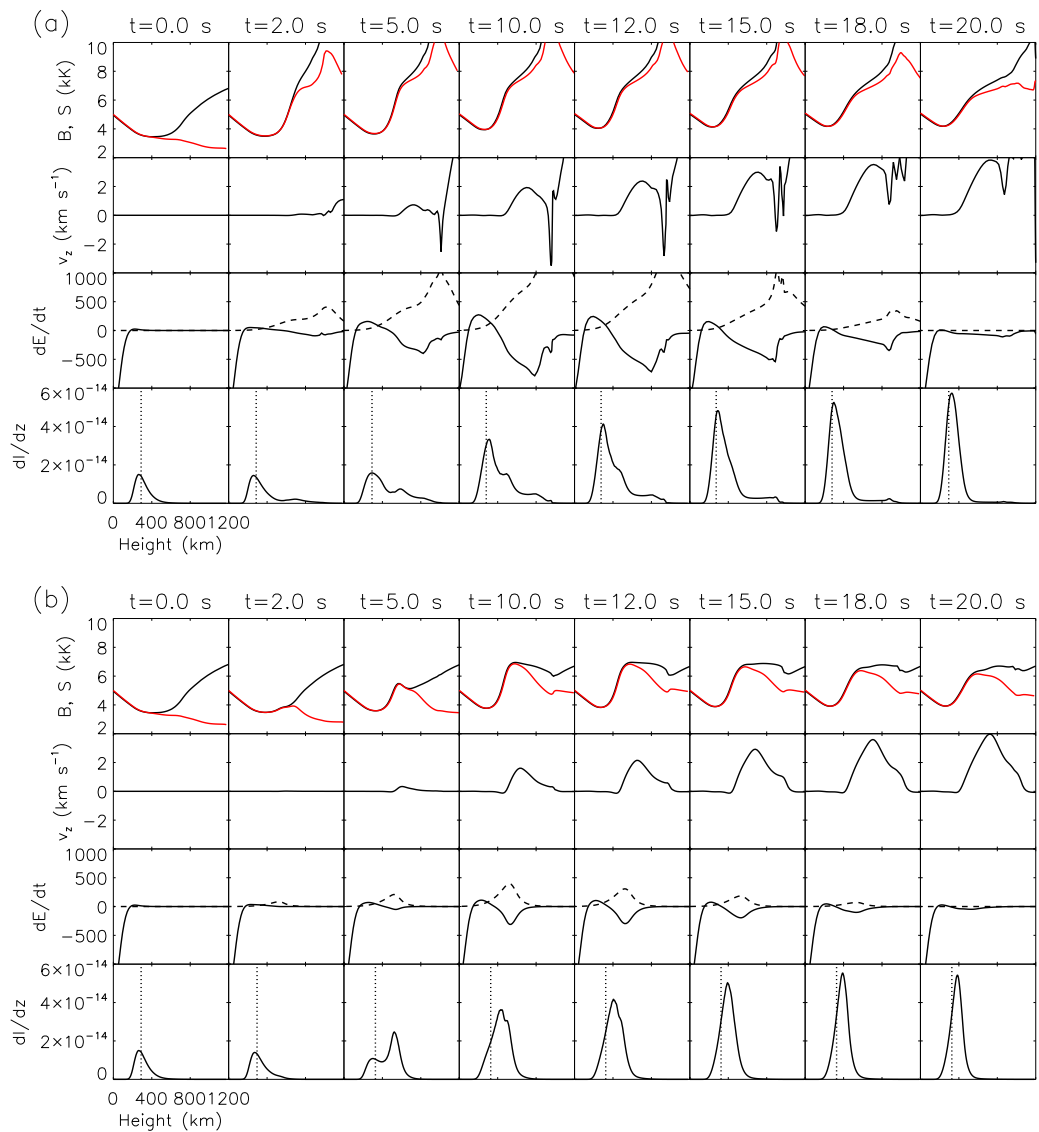}
\caption{Same as Fig.~\ref{flarea}, but for Cases FPa (a) and FPb (b).}
\label{flareb}
\end{figure}

\begin{figure}
\centering
\includegraphics[height=\textheight]{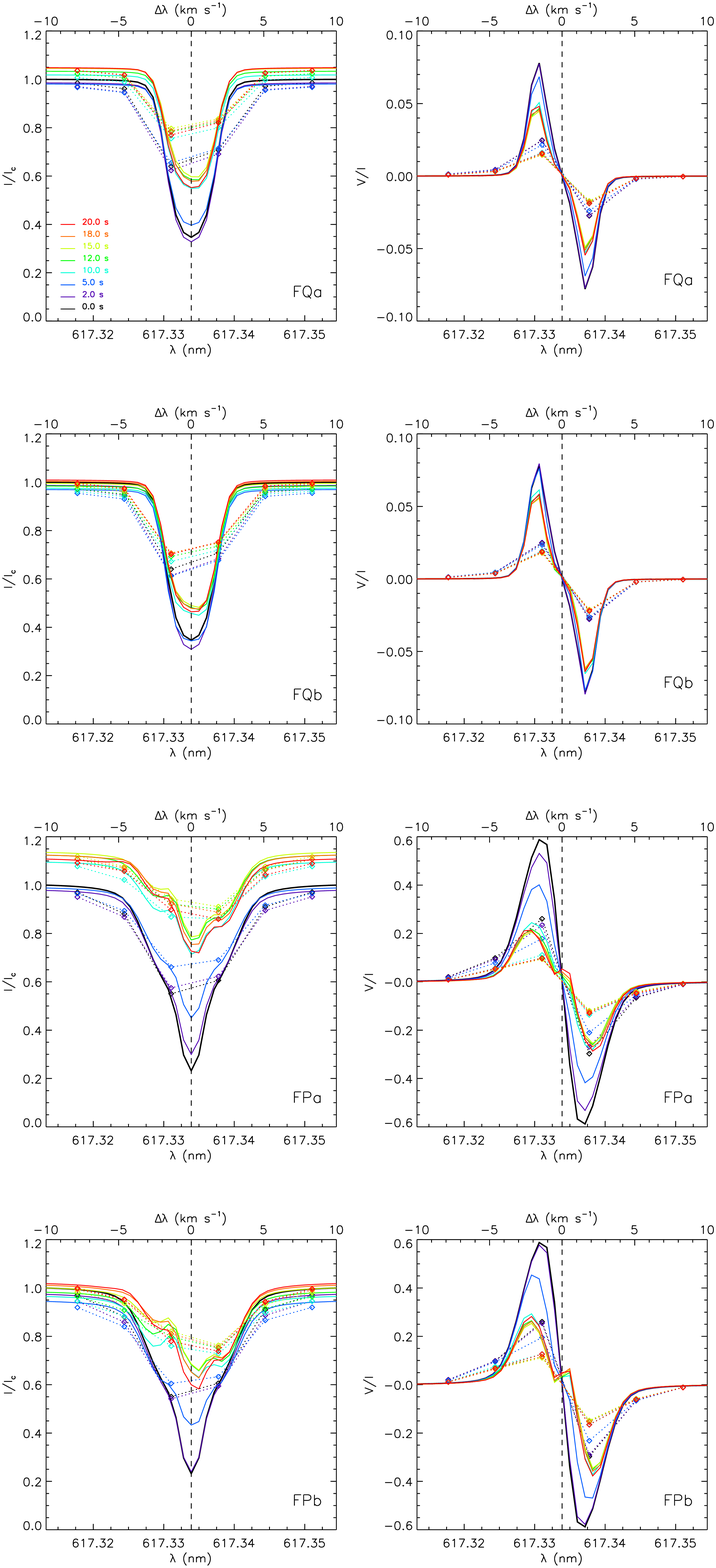}
\caption{Time evolution of the synthetic \ion{Fe}{1} 6173 \AA\ Stokes \textit{I} (left column) and \textit{V} (right column) profiles in the four simulation cases. Also shown are the simulated HMI profiles with the six tunable filters. The notations are the same as in Fig.~\ref{line}.}
\label{stokes}
\end{figure}

In cases FQa and FQb, since the magnetic field is relatively weak, the Stokes \textit{I} profiles are nearly identical to the line profiles in Fig.~\ref{line}. One can see an increase at the line center and a weak blue asymmetry. However, in cases FPa and FPb, there is a large difference in the Stokes \textit{I} profiles due to the presence of a strong magnetic field. The line width is increased here as a result of Zeeman effect. The intensity at the line center is also largely increased but no emission peak is shown here as in Fig.~\ref{line}. Yet there exist two small humps at the near wings (about $\pm 0.3$ \AA ), with a larger one at the blue wing. The simulated HMI Stokes \textit{I} profiles have a similar shape to those in Fig.~\ref{line}. For the simulated Stokes $V$ profiles, we only find a decrease in the amplitude of the lobes, but no polarity reversal, in all the simulation cases. We note that, as the HMI inversion code can automatically take into account the convolution effect by the filters, such a decrease in the magnitude of Stokes profiles does not imply a significant underestimation of the real magnetic field strength.

\section{Discussion and Summary}
In all the four cases of flare simulations, we find a change of the \ion{Fe}{1} 6173 \AA\ line in response to flare heating. Our calculations indicate that radiative backwarming and direct electron beam heating provide energy input in the photosphere and lower chromosphere, respectively; both contribute to the increase of the intensity at the line center. A higher low-energy cutoff of non-thermal electrons can increase the contribution of beam heating to the lower atmosphere and thus the line intensity. If the initial atmosphere is the quiet-Sun atmosphere, the line profile is still in absorption, but with a significant intensity increase at the line center when flare heating sets in. The formation height of the line center is still in the photosphere (around 200--300 km), but the lower chromosphere can also have some contribution to the line intensity. However, if the initial atmosphere is a cooler one like the sunspot penumbra, the temperature enhancement is larger and the line change is more obvious under the same heating condition. The line center shows an emission peak as a result of heating in the lower atmosphere. The line source function can be increased so that the contribution function peaks above the layer of optical depth unity (the line formation height before flare heating). A blue asymmetry is seen as a result of mass motions in the chromosphere.
 
The simulated Stokes \textit{I} and \textit{V} profiles can also be altered as a result of flare heating. There is a clear intensity increase at the line center of Stokes \textit{I} profiles, and the amplitude of the lobes of the Stokes \textit{V} profiles is greatly reduced, thus influencing the inverted magnetic fields.
Considering that the flare heating, in particular by electron beam bombardment as assumed in our simulations, is impulsive with always a short duration, our results provide a possible explanation that, at least some, if not all, of the magnetic transients observed in solar flares are flare-induced artifacts. 
It should be noticed that some previous observations of magnetic transients associated with flares showed similar enhancement and blue asymmetry of this line \citep{2017sun}, and sometimes an emission peak in the line center \citep{2012maurya,2017mravcova}. 
Yet we would like to point out that the HMI spectral resolution can attenuate the intensity enhancement at the line center. For example, the emission peak at the line center in cases FPa and FPb is smoothed out and not visible anymore in the simulated HMI profiles. One should also be cautious when interpreting the line asymmetries from the HMI observations, as the observed profiles would indicate a fake red asymmetry even when the original profile is symmetric. 

There has been previous calculations of the line profiles. \cite{2013harker} used semi-empirical models and reproduced this line in different emission stages, confirming that the spectral line profile can be altered by different temperature structures in the atmosphere. The role of beam heating in a flare is also investigated by calculations of this line from RADYN modeling results, which showed an enhancement at the line center \citep{2017sharykin}. They used a larger spectral index ($\delta=4$) of the electron beam and found that this line showed a redshift coinciding with the beam heating function, which is followed by a blueshift. In our results, we find the line to be generally enhanced, with an emission in the line center if the initial background atmosphere is relatively cooler like the penumbra. Besides, we only find blueshifted lines during the whole simulation time.

The fact that the response of the line to flare heating is more significant in a cooler atmosphere implies that magnetic transients should be more frequently observed in sunspot regions, which seems to be supported by observations \citep{2001kosovichev,2012maurya,2015burtseva}. However, by adopting a medium strong electron beam, we are unable to generate a full emission profile as in \cite{2013harker}, which might be possible if we increase the energy flux of the electron beam to an exceptionally high one, as implied in some recent observations \citep{2017kowalski}. In this way, there might also be a polarity reversal in the Stokes \textit{V} profiles as suggested by \cite{2013harker}.

\acknowledgments
We thank the referee for valuable suggestions that helped improve the Letter. J.H. would like to thank Zhen Li and Yang Guo for their help in parallel computing, and Viacheslav Sadykov and Jianxia Cheng for their help in numerical calculations, and Brian Harker for providing the HMI transmission profiles. We also thank Yang Liu and Alberto Sainz Dalda for helpful discussions on the magnetic field inversion from HMI data. This work was supported by NSFC under grants 11733003 and 11533005, and NKBRSF under grant 2014CB744203, and by the Research Council of Norway through its Centres of Excellence scheme, project number 262622. Y.L. is supported by CAS Pioneer Hundred Talents Program and XDA15052200.

\clearpage

\end{document}